\documentstyle[epsfig]{mn2e}{}
\begin{document}

\def\mpc{h^{-1} {\rm{Mpc}}}
\def\up{h^{-3} {\rm{Mpc^3}}}
\def\uk{h {\rm{Mpc^{-1}}}}
\def\lsim{\mathrel{\hbox{\rlap{\hbox{\lower4pt\hbox{$\sim$}}}\hbox{$<$}}}}
\def\gsim{\mathrel{\hbox{\rlap{\hbox{\lower4pt\hbox{$\sim$}}}\hbox{$>$}}}}
\def\kms {\rm{km~s^{-1}}}
\def\apj {ApJ}
\def\aj {AJ}
\def\mnras {MNRAS}
\def\aap {A\&A}

\title[Voids in the 2dFGRS]{Voids in the 2dFGRS and $\Lambda$CDM simulations: spatial and dynamical properties}
\author[Ceccarelli et al.]{
\parbox[t]{\textwidth}{Ceccarelli, L.$^1$, Padilla, N.D.$^2$, Valotto$^1$, C., Lambas, D.G.$^1$
}
\vspace*{6pt}\\
$^1$ IATE, Observatorio Astron\'omico de C\'ordoba, Argentina. \\
$^2$ Departamento de Astronom\'\i a y Astrof\'\i sica, Pontificia
     Universidad Cat\'olica, V. Mackenna 4860, Santiago 22, Chile.\\
}

\date{\today}

\maketitle

\begin{abstract}
We perform a statistical study on the distribution and dynamics of voids
in the 2dFGRS.  Our statistics are tested and calibrated using mock 2dFGRS
catalogues.  
We analyse the dynamics around voids
in mock and real 2dFGRS surveys.  
The void-galaxy cross-correlation redshift-space distortions  
show evidence of the continuing growth of voids, confirming another
prediction of the hierarchical clustering scenario. 
A non-linear outflow model can be used to provide quantitative 
estimates of the outflow velocities
around 2dFGRS voids.  These results are consistent with maximum outflows of $ 110$km/s,
$210$km/s and $270$km/s for voids of $<r_{void}>=7.5$, $12.5$ and $17.5$h$^{-1}$Mpc,
assuming a galaxy bias of $b=1$.
As an application for future surveys, our study of the mock catalogues shows
that direct measurements of the expansion of voids can be obtained using peculiar velocity data.
We find that it would also be possible to detect differences in the velocity dispersion
of galaxies in the directions parallel and perpendicular to the void walls.
\end{abstract}


\section{Introduction} 
The large scale structure in the observed galaxy distribution is characterised 
by clusters, walls, filaments and also nearly empty regions called voids 
(Vogeley, Geller \& Huchra, 1994 and references therein). 
The existence of voids has been confirmed by several surveys at a variety of 
wavelengths (Hoyle \& Vogeley, 2002), such as 
the Center for Astrophysics Survey (CfA, Vogeley et. al, 1991; 1994), 
the Southern Sky Redshift Survey (SSRS, Gazta\~naga \& Yokohama, 1993), 
the Point Source Catalogue Redshift Survey (PSCz, Hoyle \& Vogeley, 2002), 
the Infrared Astronomical Satellite (IRAS, El-Ad et. al, 1997), 
the Las Campanas Redshift Survey (LCRS, Muller et. al, 2000),
the 2degree Field Galaxy Redshift Survey (2dFGRS, Hoyle \& Vogeley, 2004; 
Croton et. al, 2004; Patiri et. al, 2005) and 
the Sloan Digital Sky Survey (SDSS, Hoyle et. al, 2005; Rojas et. al, 2005).
In all these surveys, voids have been found to occupy the greatest volumes in 
the universe.
 
Voids are found in numerical simulations as well, either identified using the 
distribution of dark-matter or haloes. 
Large scale cosmological simulations have been used to study the growth and 
evolution of voids (Regos \& Geller, 1991; Dubisnki et.al, 1993 and references 
therein).
Important progress in understanding the evolution of voids, distribution of void
sizes and void density profiles was made recently by several authors using cold
dark matter numerical simulations and mock galaxy catalogues (for instance 
Benson et al. 2003; Goldberg \& Vogeley 2004; Sheth \& van de Wiegaert 2004; 
Patiri et al. 2004; Shandarin et al. 2004; Colberg et al 2005; Padilla, Ceccarelli \& Lambas 2005 and references therein).  
Other studies have focused on the properties of halos populating the inner 
volumes of voids (Antonuccio-Degelou et al. 2002; Gottloeberg et al. 2003; 
Goldberg \& Vogeley 2004; Patiri et al. 2004; Goldberg et al. 2005).
Another widely used application of void statistics is that of probing the bias in
the galaxy distribution using large redshift surveys (Mathis \& White 2002; 
Arbabi-Bidgoli et al. 2002; Benson et al. 2003; Goldberg \& Vogeley 2004).

Void statistics have also been found to provide invaluable information on 
higher order clustering (White 1979; Fry 1986) which can be used to probe models
for galaxy clustering (Croton et.al, 2004) and void properties such as sizes, shapes 
and frequency of occurrence, and how these properties vary with galaxy type.
In general, void statistics may provide important clues on the galaxy formation
processes and can be used to place constraints on cosmological models (Peebles 
2001).
 
Sheth and van de Weygaert (2004) have developed a model for the distribution of
void sizes and evolution, and proved that surviving voids in the Universe should
be in expansion at the present time. 
Using simulations in redshift-space, Ryden (1995) found that voids show 
elongated shapes in the direction of the line of sight, a confirmation of the 
expansion of voids.
Regos \& Geller (1991) have obtain outflow velocities from simulations of the growth
of voids within the large scale structure traced by galaxies.
More recently, Padilla, Ceccarelli \& Lambas (2005) have detected the expansion
of voids, both from the redshift space distortions of void-mass and void-galaxy cross
correlation functions in numerical simulations, and 
quantitatively by measuring directly positive outflows of void shells. 
 
Based on these previous results, the central idea in this work consists on 
performing a statistical study of dynamical properties of galaxies around voids
in real and mock galaxy catalogues, to prove whether current observational 
catalogues can be used to confirm the predictions from numerical models.  
We will develop several statistical procedures which when applied to mock galaxy
catalogues will answer these questions.  
Finally, we will put the hierarchical clustering paradigm to the test once more,
by performing several void statistics using the 2dFGRS data.

This paper is organised as follows.  
We introduce the observational and synthetic catalogues in section 2.
In section 3 we describe the algorithm developed to identify voids and we apply
it to the observational and mock catalogues.
In section 4 we analise the density profiles of mock galaxies around voids in 
real and redshift space; and in section 5 we use redshift space distortions, 
peculiar velocities, and a non-linear approximation to determine properties of 
the peculiar velocity field around voids, including the amplitude of the 
expansion of voids and the dispersion of galaxies in the directions parallel and
perpendicular to the void walls.
Finally, in section 6, we present a summary of our results.\\

\section{Data}

\subsection{Observational Data: 2dFGRS }

We use the 2dF Galaxy Redshift Survey (2dFGRS) final data release (Colles et. al, 2003)
as our observational dataset.  
The 2dFGRS is one of the widest spectroscopic data sets available to date, and 
contains spectroscopic redshifts for approximately $230,000$ galaxies.
The source catalogue for the 2dFGRS is a revised and extended version of the APM
galaxy catalogue from which a set of target galaxies, characterised by 
extinction-corrected magnitudes brighter than b$_J$=19.45, was selected for the
construction of the 2dFGRS. 
The main survey regions are two declination strips, plus 99 randomly located 
fields.
Our results are obtained using the two strips described below.
One strip is located near the southern Galactic pole, spanning 
approximately $80$x$15$ degrees; the other strip is centred in the Northern 
Galactic pole spanning $75$x$10$ degrees along the celestial equator. 
The survey covers a total of $2,000$ square degrees and has a median depth of 
$z=0.11$.

The completeness of the survey varies with position due to unobserved fields, 
observed objects with poor spectra, and objects that could not be observed due 
to either fiber collision constraints or broken fibers. 
The completeness 
is due in the range 
 [0, 1] .
To match the angular selection function in the construction of random catalogues
we use the software developed by the 2dFGRS team and distributed as part of the
data release
that, for any coordinates $(\alpha, \delta)$, gives the expected probability of
galaxy being contained in the catalogue.

\subsection{Synthetic Data: Mock 2dFGRS}

We use a $\Lambda$CDM numerical simulation populated with GALFORM
semi-analytic galaxies as our starting point for creating mock 2dFGRS
catalogues. The numerical simulation follows the evolution of $250^3$
particles in a ``Concordance" Cosmology with parameters in line with the
results from the Wilkinson Microwave Anisotropy Map (WMAP, Spergel et al.
2003) on the temperature fluctuations of the Cosmic Microwave Background,
and clustering measurements from the 2dFGRS and Sloan Digital Sky Survey
(Abazajian et al. 2004).  Specifically, the box side measures
$250$h$^{-1}$Mpc, the matter density parameter corresponds to $\Omega_m=
0.3$, the value of the vacuum density parameter is $\Omega_v=0.7$, the
Hubble constant, $H_0=h 100$s$^{-1}$km Mpc$^-1$, with $h= 0.7$, and the
primordial power-spectrum slope is $ns= 0.97$. The present-day amplitude
of fluctuations in spheres of $8$h$^{-1}$Mpc is set to $\sigma_8=0.8$.  
The resulting mass per dark-matter particle is $M_{min}= 1.05 \times
10^{10}$h$^{-1}$M$_{\odot}$.  The semi-analytic galaxies populating this
numerical simulation were kindly provided by the Durham group (private
communication). We briefly explain the procedure followed to this end:
dark-matter haloes of at least $10$ dark-matter particles are first
identified in the numerical simulation using a Friends-of-Friends
algorithm with a linking length $b=0.2$.  The resulting haloes are
characterized by a minimum mass of $M_{min}= 1.05 \times
10^{11}$h$^{-1}$M$_{\odot}$, a median mass of $M_{min}= 2 \times
10^{11}$h$^{-1}$M$_{\odot}$, and a maximum mass of $M_{min}= 2 \times
15^{11}$h$^{-1}$M$_{\odot}$. These haloes are assigned galaxies using the
GALFORM code (Cole et al. 2000, Baugh et al. 2005). We use these
semi-analytic galaxies to construct mock 2dFGRS surveys by first
positioning an observer in a random position within the numerical
simulation box, and then reproducing the selection function and angular
mask of the 2dFGRS from this position.  This results in a catalogue of
galaxies of similar properties and observational biases to those of the
real catalogue.  This mock catalogue will be used in this work to
calibrate our statistical methods, to interpret the data, and to detect
any systematic biases in our procedures.  In order to do this, we will
treat the mock catalogue in exactly the same way as the real data. We
start our analyses by defining the same galaxy subsamples for real and
mock data in the next subsection.

\subsection{Galaxy Samples}
Since the 2dFGRS is an apparent magnitude limited survey, the faintest galaxies
are only registered at small distances while the brightest galaxies cover the 
whole catalogue. 
This results in a luminosity bias which can be overcome by constructing volume 
limited samples. 

For a given maximum redshift $z_{lim}$ the catalogue is complete for galaxies 
brighter than 
   \begin{equation}
   B_{lim}=b_{lim}-25-5*log(c*z_{lim})+5*log(H),
   \label{eq:blim}
   \end{equation}
where $b_{lim}$ is the flux completeness limit and $H$ is the Hubble constant.
This magnitude limit varies from point to point in the catalogue, with a 
brightest magnitude limit cut of $b<18.9$ (the whole catalogue is complete above
this cut).

We construct several volume limited samples with galaxies brighter than 
$B_{lim}$ and redshifts $z<z_{lim}$.
The aim to use several samples is is to compare results, differences in 
$z_{lim}$ are related with differences in size of the identified voids and 
differences in $B_{lim}$ are useful to check the underdensity stability when 
using different average galaxy densities.
We choose 3 limits in redshift and 2 luminosity limits.  
As a result we obtain 12 samples in total for both, the 2dFGRS and mock catalogues.
 
The parameters $z_{lim}$ and $B_{lim}$ used to restrict the samples are shown in
the second and third columns of Table 1.  
The remaining columns of the table show the resulting number of galaxies in each
subsample for the NGP and SGP regions and real and mock catalogues.
The values of $B_{lim}$ chosen for each subsample satisfy equation \ref{eq:blim}
for the quoted values of $z_{lim}$
and $b_{lim}=18.9$ for $S121$, $S151$ and $S171$, 
$b_{lim}=19.6$ for $S122$, $S152$ and $S172$. 

    \begin{table}
    \begin{tabular}{ccccccc}
    \hline
    \hline
    \noalign{\vglue 0.2em}
    sample  & $z_{lim}$ & $B_{lim}$ & $N^{mock}_{ngp}$ & $N^{mock}_{sgp}$ & $N^{2df}_{ngp}$ & $N^{2df}_{sgp}$\\
    \noalign{\vglue 0.2em}
    \hline
    \noalign{\vglue 0.2em}
    $S121$ &  0.12  &  -18.88 & 20079 & 29173 & 16102 & 21086 \\
    $S122$ &  0.12  &  -19.58 & 10162 & 14694 & 5382  & 7231\\
    $S151$ &  0.15  &  -19.36 & 23235 & 32999 & 13816 & 18128 \\
    $S152$ &  0.15  &  -20.07 & 9610  & 13706 & 3092  & 4239\\
    $S171$ &  0.17  &  -19.64 & 22529 & 33191 & 10349 & 14660 \\
    $S172$ &  0.17  &  -20.34 & 8260  & 11984 & 1668  & 2615 \\
    \noalign{\vglue 0.2em}
    \hline
    \hline
    \end{tabular}\label{table:glxsamples}
    \caption{Definition of galaxy subsamples in the NGP and SGP regions of mock
     and real 2dFGRS catalogues.}
    \end{table}

As can be seen in table 1 the number and number density of galaxies diminish 
when the sample is restricted to the brighter galaxies and to higher redshifts.
This will have an impact on the identification of voids.
Different samples will be the most appropriate to search for voids of different
sizes according to the sample volume. 
For instance, greater volumes are better suited to find a statistically relevant
number of large voids.  
However, in such cases the density of galaxies is low and it becomes difficult 
to find small voids.
In the opposite case, a smaller volume with a denser population of galaxies 
allows us to identify small voids, but only includes a handful of large voids.  
We take this into account when choosing a subsample of galaxies to identify 
voids of a given radius.

\section{Voids in galaxy catalogues}

\subsection{Void finding algorithm}

We construct an algorithm to identify voids in galaxy catalogues.  
In order to develop this algorithm we assume that voids are approximately 
spherical. 
Our assumption is based on the approximate shapes of empty regions in 
numerical simulations (Padilla, Ceccarelli \& Lambas 2005).
The 'a priori' knowledge of the void shape suggests the way in that the 
algorithm will work, that is by searching empty spherical regions. 
We define a void as the largest spherical volume within which the matter density
is below a critical value. 
According to this definition, two concentric spheres of different radii can not
be different voids.
Similarly, two or more superimposed empty spheres are not considered as individual 
voids. 

The algorithm consists of the following steps: 
First, we set a large number of random positions for void centre candidates. 
These random positions are distributed throughout the catalogue.
Notice that the efficiency of the algorithm is partly determined by the density
of random centres.  Therefore, we choose the number of candidate centres such
that the mean distance between centres is smaller than the mean 
distance between galaxies.  
Moreover, we always use the largest possible
number of centres within our computational limits.
We also check that the relation 
$V_{survey}/V_{sphere}$ remains large enough for all the void sizes considered.
In all random positions we consider spheres with radii ranging from r$_{min}$ to
r$_{max}$ 
(every random position is the centre of a sphere whose 
radius increases from r$_{min}$ to r$_{max}$).
Given that by definition r$_{min}$ is the minimum possible void radius, we 
carefully choose an appropriate value for this parameter.
The density contrast ($\delta_{gx}$) is measured for every sphere. 
At this point the algorithm computes the density within each sphere as a function
of radius
and this value is compared to the maximum allowed value, $\delta_{max}$ (this
allows several candidates per centre, all with different radii).  
All the spheres satisfying the condition $\delta_{gx} < \delta_{max}$ are 
considered underdense spheres and are selected as void candidates.  
At this point it is still possible that many void candidates share the same 
centre while having different radii. 
In such a case, we always keep the largest sphere.
Finally, once the underdense spheres have been selected, we decide which 
underdense spheres are voids.
In order to do this we consider three cases:
If there is a sphere without any superpositions, this sphere is a void.
If there is a sphere within a larger sphere, the smaller candidate is removed.  
If there are two or more superimposed voids, all of them are removed and the 
only one considered as a void is the largest underdense sphere, which should
contain the others.

\subsection{Voids in mock and 2dFGRS galaxy Subsamples}
We apply the void finding algorithm to the mock and real 2dFGRS samples defined
in the previous section.  
In this work we only identify voids using galaxy positions in redshift-space. 
Padilla, Ceccarelli \& Lambas (2005) showed that the resulting statistics change
only slightly when using voids identified in redshift-space with respect to 
using voids identified in real-space.
When calculating the density contrast required in our algorithm we take into 
account the survey incompleteness. 
For this aim we use the mask software developed by the 2dFGRS team.

We find voids with radii ranging from $r_{min}=5$ Mpc h$^{-1}$ to $r_{max}=20$ Mpc h$^{-1}$
corresponding to the range allowed by the void size constraints suggested by the void 
probability function in the 2dFGRS (Croton et al. 2004; Hoyle \& Vogeley 2004). 
Also, according to results obtained by Hoyle \& Vogeley (2004)
void-radii are usually smaller than 25 Mpc/h. 
Another important reason behind the chosen value of $r_{min}$ is given by the
galaxy density in our 2dFGRS sub-samples.
As we are interested in extremely underdense regions we set $\delta_{max}=-0.9$.
This value is in agreement with the mean density contrast of voids identified
by Hoyle \& Vogeley (2004).

In table 2 we show the total number of voids obtained in each sample.
Note that the brightest samples (S083, S103, S123, 
S153 and S173) contain a small number of voids with respect to
samples constructed using the same $z_{lim}$ and fainter galaxies.  

    \begin{table}
    \begin{tabular}{ccccc}
    \hline
    \hline
    \noalign{\vglue 0.2em}
    &mock north&mock south&2dF north&2dF south\\
    sample  & $N_{ngp}$ & $N_{sgp}$ & $N_{ngp}$ & $N_{sgp}$\\
    \noalign{\vglue 0.2em}
    \hline
    \noalign{\vglue 0.2em}
    $S121$ &  70  &  88 &  73 & 53 \\
    $S122$ &  49  &  61 &  40 & 29 \\
    $S151$ &  86  & 135 &  64 & 78 \\
    $S152$ &  39  &  71 &  25 & 34 \\
    $S171$ & 109  & 146 &  65 & 76 \\
    $S172$ &  54  &  68 &   2 &  9 \\
    \noalign{\vglue 0.2em}
    \hline
    \hline
    \end{tabular}\label{table:fits}
    \caption{Number of voids identified in the mock and real 2dFGRS, in the 
             NGP and SGP regions. }
    \end{table}

Taking into account the different sizes of voids identified in each sample, a 
more detailed analysis of our samples of voids reveals that:
\begin{itemize}
\item In the smallest samples (S121, S122) we do not find a significant 
number of voids with radii larger than $10$h$^{-1}$Mpc. 
However, specially in the most numerous sample (S121), we find a good number of 
voids with radii smaller than about $7$h$^{-1}$Mpc.   
\item Samples with the smallest volumes (S121 and S122) 
are more suitable for searching voids of radii between 5 and 10 Mpc/h rather 
than larger voids.   
\item The intermediate volume samples (S151, S152) give us the maximum 
number of voids with radii of $12$, $13$, $14$ and $15$h$^{-1}$Mpc.
These samples are better suited for searching voids of radii between 10 and 15 
Mpc/h rather than either smaller or larger voids.   
\item In the larger volume samples (S171 and S172) we find the largest 
number of large voids ($r_{voids}> 15 Mpc/h$) whereas we do not find any small 
voids.
\end{itemize}

According to this analysis the use of small volume samples is suitable
for a detailed study of small voids.
The final selection of samples includes, 
\begin{itemize}
\item{S121}, with z$<$0.12 and B$<$-18.88, $5\leq r_{void}/Mpc/h\leq10$     
\item{S151}, with z$<$0.15 and B$<$-19.36, $10<r_{void}/Mpc/h\leq15$     
\item{S171}, with z$<$0.17 and B$<$-18.94, $15<r_{void}/Mpc/h\leq20$     
\end{itemize}
 
    \begin{figure}
    \epsfxsize=0.5\textwidth
    \hspace*{-0.5cm} \centerline{\epsffile{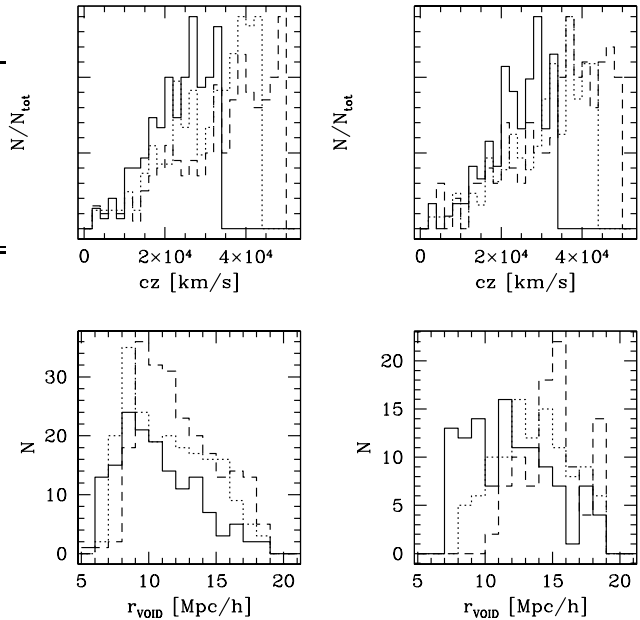}}
    \caption{Radial velocity (upper panels) and void radii (lower panels) 
             histograms for voids identified in the (right panels) and 
             2dFGRS catalogues (left panels). 
             Different lines correspond to different samples,
             solid line for sample S121 , 	
             dotted line for sample S151 and 	
             dashed line for sample S171.
             } 
    \label{fig:hzrm}
    \end{figure}

In figure \ref{fig:hzrm} we show the redshift distributions of void centres 
(upper panels) and distributions of void radii (lower panels) for voids in 
samples C121 (solid lines), C151 (dotted lines), and C171 (dashed lines), which
correspond to $z < 0.12$, $z < 0.15$ and $z < 0.17$ respectively, for the 2dFGRS
mock (left panels) and real (right panels) catalogues.  
As it can be seen in figure \ref{fig:hzrm}, the distributions are
similar in both mock and 2dFGRS
catalogues, indicating that the general characteristics of
voids are comparable and therefore the mock catalogue is suited to make
appropriate comparisons to the real data.

\section{Spatial distribution of galaxies around voids}

\subsection{Galaxy density profiles}

In this section we study the density profiles of galaxies around voids.
We are specially interested in the analysis of distortions in the profiles 
caused by the use of galaxy positions in redshift-space.
We also test the quality of the voids identified in this work by examining the 
density profiles as a function of distance to the void centre in the mock 
catalogues.  
In all figures, errors are calculated using the $10$ jackknife subsamples of 
voids (see Padilla, Ceccarelli \& Lambas 2005, for more details on error 
measurements).
In order to apply this method for the estimate of errors we follow
the usual
procedure,
whereby
we calculate the density profile for $10$ sub-catalogues of galaxies
extracted from the sample of galaxies under analysis.  
Each one of these sub-catalogues consists of the actual sample with one-tenth of its volume removed; 
this procedure excludes approximately one-tenth of the galaxies and
each member is only removed once.  
The variance of the results of the true sample is that of
the results derived from the $10$ sub-catalogues.

The analysis of void profiles resulting from different samples generally shows 
extremely empty voids, with approximately zero density of galaxies near the void
centres. 
Densities smaller than 5$\%$ the median are common in the void interiors and
the galaxy density abruptly increases at distances of about the void radius.
This is a good indication that we accurately find the boundaries of voids.

   \begin{figure}
   \epsfxsize=0.5\textwidth
   \hspace*{-0.5cm} \centerline{\epsffile{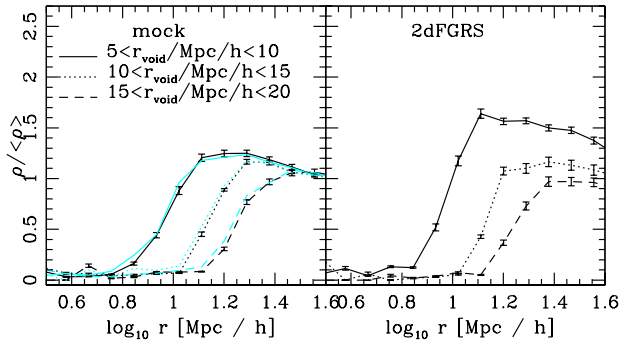}}
   \vspace*{-3.5cm}
   \caption{Galaxy density profiles of mock (left) and 2dFGRS 
   (right) galaxies as a function of distance to the void centre.
   For the mock catalogue, we show results in real- and redshift-space 
   in gray (cyan) and black lines, respectively.  
   Different line types are associated with different subsamples and indicate 
   different void radii: solid lines correspond to $5\leq r_{void}/Mpc/h\leq 10$
   and sample S121, dotted lines correspond to $10<r_{void}/Mpc/h\leq 15$ and 
   S151, and dashed lines correspond to $15<r_{void}/Mpc/h\leq 20$ and S171.
   The sample used for the void identification is the same used to trace the 
   profile.} 
   \label{fig:denm}
   \end{figure}

   \begin{figure}
   \epsfxsize=0.5\textwidth
   \hspace*{-0.5cm} \centerline{\epsffile{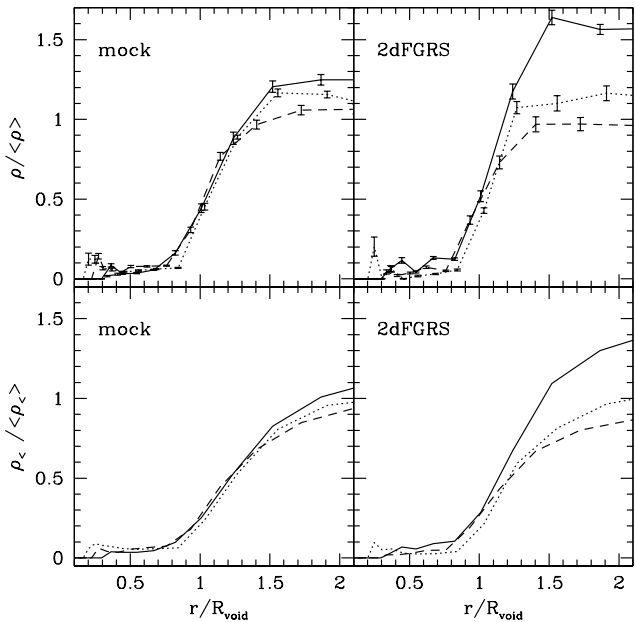}}
   \caption{Galaxy density profiles as a function  of normalized distance 
   to the void center. Upper panels: differential density profiles.
   Lower panels: cumulative galaxy density profiles.  
   Galaxy samples and line types are the same as in
   figure \ref{fig:denm}}  
   \label{fig:dencomp}
   \end{figure}

   \begin{figure}
   \epsfxsize=0.5\textwidth
   \hspace*{-0.5cm} \centerline{\epsffile{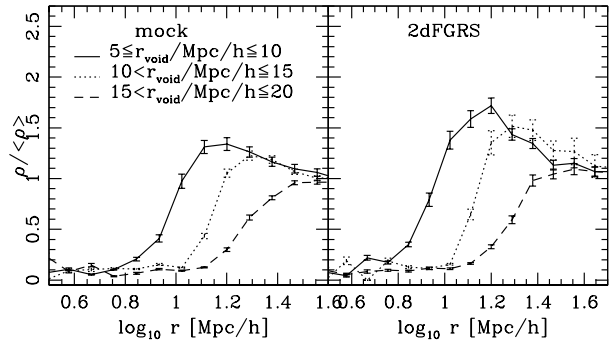}}
   \vspace*{-3.5cm}
   \caption{Density profiles of mock (left) and real (right) galaxies as a 
   function of distance to the void centre.  
   The density profile is traced by fainter galaxies than those used to identify
   the voids; for $5\leq r_{void}/Mpc/h\leq 10$ (solid line) voids are from S122
   and profiles from S121, for $10<r_{void}/Mpc/h\leq 15$ (dotted line) voids 
   are from S152 and profiles from S151, and for $15<r_{void}/Mpc/h\leq 20$ 
   (dashed line) voids are from S172 and profiles from S171.} 
   \label{fig:dencross}
   \end{figure}

In figure \ref{fig:denm} we show the density profiles for voids in the mock 
catalogue in real and redshift space (left panels) and for the 2dFGRS survey,
only in redshift space (right panels).  
Each figure corresponds to a different sample in the mock and 2dFGRS catalogues
as is indicated.  
We analise the density profiles as a function of the distance to the void 
centre.
As can be seen in this figure the profiles show the same behavior 
regardless of the void radius.  
However, if we analise in detail the density profiles as a function of void 
radius we can detect some differences. 
The shape of profiles of the smaller voids reach the higher values at a 
distance of $\simeq 1$ void radius. The largest voids only reach
profile values of $\simeq 1$ at the same distance of $1$ void radius.
This behavior is consistent with the results from semi-analytic galaxies in the
simulation box (Padilla, Ceccarelli \& Lambas, 2005). 
Note that the samples used in this figure correspond to different absolute
magnitude limits for different void sizes as explained in section 2.3. 

In the case of the mock catalogue, we consider the density profiles in real- and
redshift-space ($\delta_{rs}$ and $\delta_{zs}$ respectively) and show them in 
the left panel of figure \ref{fig:denm}.
As can be seen in the figure, the profiles show the same shape in real and 
redshift space, although we can appreciate a difference at a distance of the 
order of the void radius where the redshift distortion in the void shape 
introduces an overestimation of the void radius in redshift space.
This redshift distortion corresponds to
the expansion of the void shells, which look elongated in the direction of
the line of sight (Ryden, 1995).  This same effect may be responsible for
the differences in redshift- and real-space profiles at distances smaller
than the void radius.
The density profile in real and redshift space tends to be the same at 
distances greater than the void radius.
In any case, from the comparison of the relative sizes of redshift-space distortion
effects and the size of error-bars in the density profiles we find that this 
systematic effect is small and therefore we do not correct for it.

In order to compare with previous works we calculate the cumulative density profiles
as a function of normalised distance to the void centre, and show the
results in the lower panels of figure \ref{fig:dencomp}.
The corresponding differential density profiles are shown for comparison
in the lower panels of figure \ref{fig:dencomp}.
As it can be seen in the lower panels, our cumulative densities 
converge to the average density at distances of about 2 void radius whereas 
the void profiles by Hoyle \& Vogeley 
(2004) remain lower than the mean density. 
The cumulative density slope at the void shell in Hoyle \& Vogeley (2004) is 
lower than ours in spite of 
being selected to have the same galaxy density inside the voids. 
This difference could be a consequence of the void identification algorithm. 
We also notice that the profiles shown in the  upper panels of 
figure \ref{fig:dencomp}, are similar to the results by Benson et al (2003), 
which were obtained from dark matter and galaxies in semianalytic simulations.   

We now focuse on fainter galaxies inside voids.  In order to do this, we
calculate the void density profiles using fainter galaxies than those used
to identify the voids.  Figure \ref{fig:dencross} shows the results.
This analysis is performed in order to investigate 
the possible existence of a faint galaxy 
population inside voids and also to test variations of the galaxy population in 
voids, and its consequences on void identification and dynamics.
As it can be seen in figure \ref{fig:dencross} 
there are almost no faint galaxies 
filling the void volumes.
We can therefore assume that the voids are indeed virtually empty, and therefore
the density profile obtained using a complete sample of galaxies can approximate
the mass density profile (consistent with semi-analytic simulation results from
Padilla, Ceccarelli \& Lambas 2005).
This assumption has important dynamical consequences as will be seen below.

\section{Dynamical properties of galaxies in voids}

We analyse the dynamics of galaxies around voids following two different 
approaches.
The first is a qualitative approach, where we study the shape of the isocontours
of the void-galaxy cross-correlation function measured in the directions 
parallel and perpendicular to the line of sight. 
From this study we can infer characteristics of the peculiar velocity field. 
The second approach is a quantitative measurement of galaxy outflows or void
expansion, which can be derived in two ways; 
i) the first measurement of the void expansion can be achieved by applying a 
simple, non-linear theory approximation that relates the peculiar velocity 
outflows to the void-galaxy cross-correlation function; this method can be 
applied to the mock and real 2dFGRS data.
ii) The second method is only possible in the mock catalogue due to the current
observational limitations, where we directly calculate the outflow amplitudes 
from peculiar velocities.
The use of peculiar velocities allows a direct, quantitative measurement of the outflow
of galaxies from voids, which can be compared to predictions from non-linear 
theory approximation and results from the numerical simulation.
We also test whether a mock peculiar velocity survey can detect subtle 
properties of the dynamics of galaxies around voids, such as a larger velocity 
dispersion in the direction of the void walls.

\subsection{Dynamics using redshift space distortions}

We have performed a study of the void-galaxy cross correlation function,
that is the correlation function between void centres and galaxies,
as a function of the directions parallel and perpendicular to line of sight, 
$\pi$ and $\sigma$ respectively.

The assumption that our Universe is isotropic leads us to expect that the 
correlation function should also be isotropic. 
However, if we measure the correlation function in redshift-space instead of in
real-space, anisotropies appear.
This deviations from isotropy are known as redshift-space distortions.
These distortions are due to peculiar velocities, and therefore can provide
information on the dynamical properties of galaxies around voids. 
$\xi (\sigma,\pi)$ is calculated using the estimator
   \begin{equation}
   \xi (\sigma,\pi)= DD(\Delta \sigma,\Delta \pi)/DR(\Delta \sigma,\Delta \pi)-1
   \end{equation}
where $DD$ is the number of galaxy and void-centre pairs in the range of 
distances ($\Delta \sigma, \Delta \pi$), and $DR$ is the number of void-random galaxy
pairs in the same range of distances 

\begin{figure*}
   \epsfxsize=1.2\textwidth
   \hspace*{-0.5cm} \centerline{\epsffile{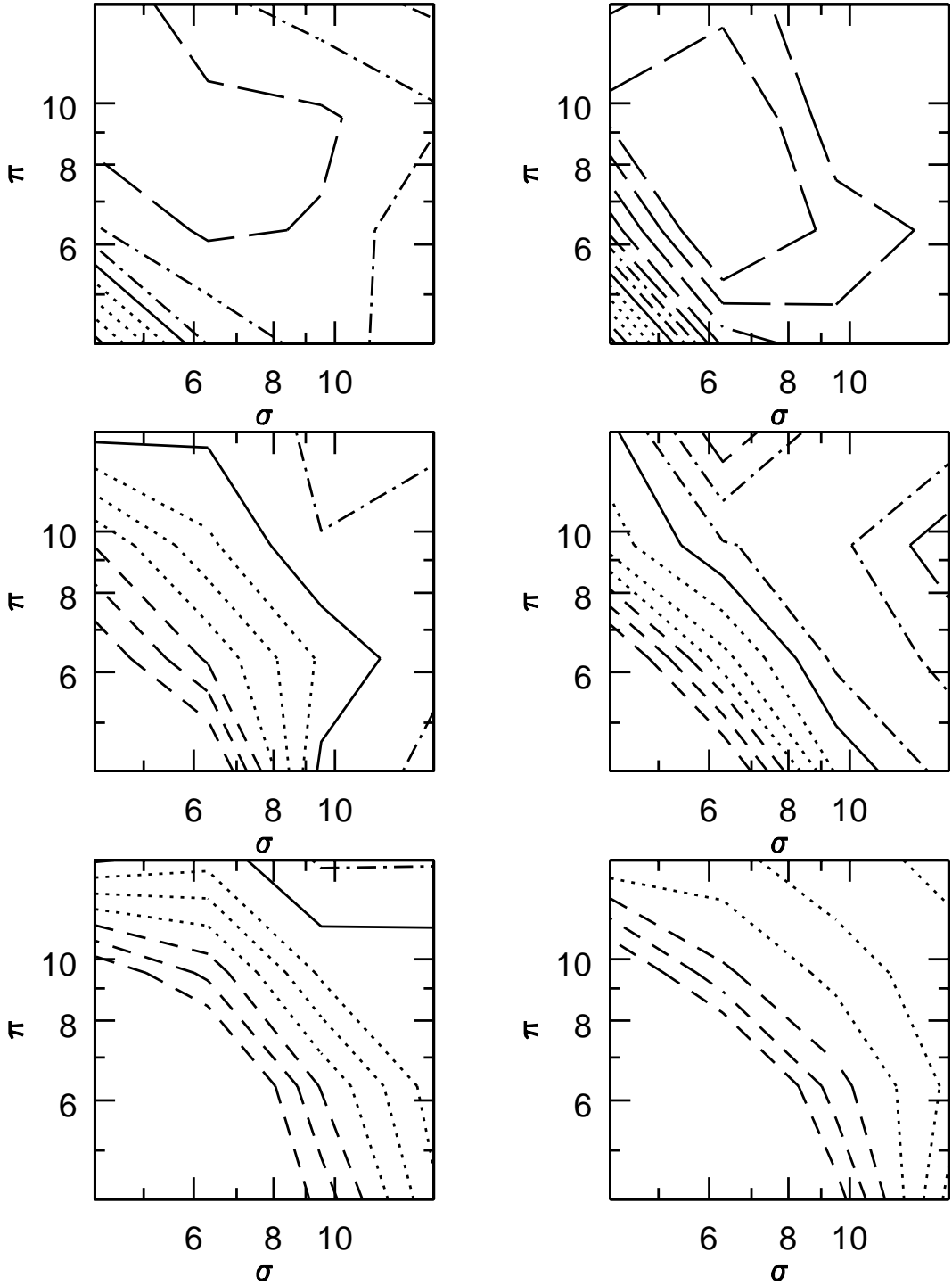}
   }
   \caption{Redshift space void-galaxy correlation function $\xi(\sigma,\pi)$ 
   estimated from mock (right panels) and real 2dFGRS (left panels) catalogues.
   $\sigma$ and $\pi$ represent separations parallel and perpendicular to the 
   line of sight. 
   Each panel corresponds to different ranges of void radii and are obtained 
   from different samples; top: $5\leq r_{void}/Mpc/h\leq 10$ and S121, middle:
   $10< r_{void}/Mpc/h\leq 15$ and S151, bottom: $15< r_{void}/Mpc/h\leq 20$ and
   S171. Dashed lines represent $\xi=-0.6, -0.5, -0.4$, 
   dotted lines represent $\xi=-0.3, -0.2, -0.1$, 
   solid line represents the iso-$\xi$ contour corresponding to $\xi=0.$,
   dot-dashed lines represent $\xi=0.1, 0.2$ and short-dashed-long-dashed lines represent
    $\xi=0.3, 0.4, 0.5, 0.6, 0.7$.
    }
   \label{fig:rsdmc1}
   \end{figure*}

In a hierarchical scenario of structure formation, the objects in underdense 
regions tend to be accreted toward more massive systems. 
According to this, the galaxies inside voids and in void shells, would be 
attracted to the matter outside of the void, producing a large scale streaming 
motion in the form of an outflow from the void centre or a void expansion.
The outflow motion would be seen in the distortion pattern of the z-space 
correlation function as an elongation in the direction of the line of sight 
($\pi$ direction).  
The upper panels of figure \ref{fig:rsdmc1} show the contour lines for the 
samples S121 from the mock (left panels) and 2dFGRS (right panels) catalogues, 
for small void radii ($5\leq r_{void}/Mpc/h \leq 10$). 
The two panels in the middle of figure \ref{fig:rsdmc1} correspond to 
intermediate void sizes $10< r_{void}/Mpc/h \leq 15$ obtained from sample S151.
The bottom of figure \ref{fig:rsdmc1} shows the distortions for the largest 
voids ($15<r_{void}/Mpc/h \leq 20$) obtained using sample S171.
As is mentioned in section 3, the different samples considered in the upper, 
middle and lower panels of the figure have been chosen to have the most 
convenient number of voids for a high statistical significance and to reduce the
noise in the results.  
The estimated jackknife uncertainties in $\xi(\sigma,\pi)$ are $\simeq10\%$ of the value of $\xi$.
There is a clear elongated distortion pattern in all panels of figure \ref{fig:rsdmc1} 
which is consistent with an outflow motion. 
The fact that the 2dFGRS galaxies show such an elongation around voids 
represents a further confirmation of the hierarchical clustering scenario of 
structure formation.
However, the noisy results restrict the quantitative applications of this 
method.  Therefore, a different approach should be followed in order to quantify 
the expansion of voids in the Universe.

\subsection{Quantitative measures of galaxy outflows around voids}

In order to study the statistical characteristics of the peculiar velocity field
around voids we estimate the mean amplitude of the velocity field as a function
of the distance to the void centre.  
We place our reference system at the void centres so that a negative velocity 
indicates an outflow motion (and a positive velocity indicates an infall onto 
the void centre).
We adopt two approaches, 
an indirect derivation of the velocity field from a theoretical model,
and 
a direct measure of the peculiar velocity amplitude. 
The first method, requires many assumptions about the adopted cosmological model
in order to derive the velocity field.  
However, this method works satisfactorily even when assuming a simple, 
non-linear approximation.
The second method consists of the calculation of the velocity field amplitude
using galaxy peculiar velocities. 
Even though this is the most simple and natural approach, it has the 
disadvantage of being only applicable to mock catalogues.

\subsubsection{Non-Linear Model for the Outflow}

We apply a non-linear theory model by Croft, Dalton \& Efstathiou (1999) to 
describe the dynamical behavior of particles around underdense regions.
We choose to use a non-linear model in favor of a linear approximation since in
the inner regions of voids, the overdensity is significantly different from zero
and therefore, the linear theory condition $|\delta|<<1$ is not satisfied.
The outflow model, which is analogous to the infall model for overdensities, 
assumes spherical symmetry.     
When analyzing both, mock catalogues and real data, we assume the concordance 
$\Lambda$CDM cosmology consistent with results from WMAP (Spergel et al. 2003).
As the overdensity within regions around voids is $\sim -1$, we can expect the 
non-linear approximation to describe the velocity field accurately,
   \begin{equation}
   V_{inf}=-\frac13\Omega^{0.6} H_0 r \delta(r)/(1+\delta(r))^{0.25} 
   \end{equation}
where $\delta(r)$ is the matter density contrast within a sphere of radius $r$.
We apply the outflow  model to predict the amplitude 
of peculiar velocities. The model uses the measured values void-galaxy
cross correlation function in redshift and real space ($\xi(s)$ and $\xi(r)$, respectively)
to calculate the galaxy density contrast $\delta_{gx}(r)$(Croft et. al, 1999);
the subindex $gx$ indicates that we can not measure
the void-mass cross-correlation function directly.  Therefore,
the infall velocities obtained from this formalism
actually correspond to
\begin{equation}
 V_{inf}^{gx}\simeq b V_{inf}
\end{equation}
valid where $|\delta|<<1$ (that is, at the void boundaries or further away).
$b$ is the unknown bias factor relating the mass and galaxy
density contrasts around voids, which in this work is assumed to be $b=1$
(see for instance Verde et al. 2002).

We first apply this formalism to the void-galaxy cross correlation function
obtained from the mock catalogue and show the predicted peculiar velocities in 
gray lines in figure \ref{fig:fn2df}, where each line-type corresponds to different 
void radii (same ranges in black lines).

We then apply the outflow model to the 2dFGRS catalogue. We use the redshift space 
correlations obtained for the observational catalogue
assuming that our results will not change significantly when using
redshift-space correlations as opposed to real-space data.
The predicted velocities are shown in black lines 
in figure \ref{fig:fn2df}. For comparison, we also show 
the predicted velocities for the mock catalogue in gray lines. 
The maximum outflow amplitudes for the 2dFGRS, and distances where this 
maximum occurs, are in excellent agreement with the 
predictions for the mock catalogue. 
Regardless of the void size, 
both curves exhibit very similar amplitudes and shapes, specially
at distances of the order of one void radius,

It is very reassuring to find 
the same velocity outflows in mock and observational 
catalogues.  This suggests we are obtaining a reliable estimate for the velocity of void 
shells in the 2dFGRS. Only at large separations from the void centres 
the outflow velocities became considerably larger in the 2dFGRS. It remains to 
be confirmed whether this result is to be expected from the data or 
simply produced by systematic effects perhaps induced by the use of
redshift-space correlations.

\subsubsection{Galaxy outflows from mock peculiar velocities }

Studies on the outflow of void shells have been performed 
mainly in numerical simulations and semi-analytical galaxies by several authors. For instance, 
Regos \& Geller have studied the evolution of voids in numerical simulations obtaining
the peculiar streaming velocities of void walls
(Regos \& Geller 1991).
Dubinski et al. (1993) have analised the peculiar velocity field surrounding voids in simulations. 
Another related work was performed by Seth \& van de Weygaert (2004) 
who studied the void size evolution in simulations.

We compute the amplitude of the velocity field 
around voids using the line-of-sight component of galaxy
peculiar velocities in mock 2dFGRS catalogues, assuming a radial, spherically symmetric 
outflow motion.
Note that as the real 2dFGRS survey contains no data regarding peculiar
velocities, this method can not be applied to this particular catalogue.  Therefore,
all the analyses in this subsection are restricted to
the mock catalogues.

\begin{figure}
\epsfxsize=0.5\textwidth
\hspace*{-0.5cm} \centerline{\epsffile{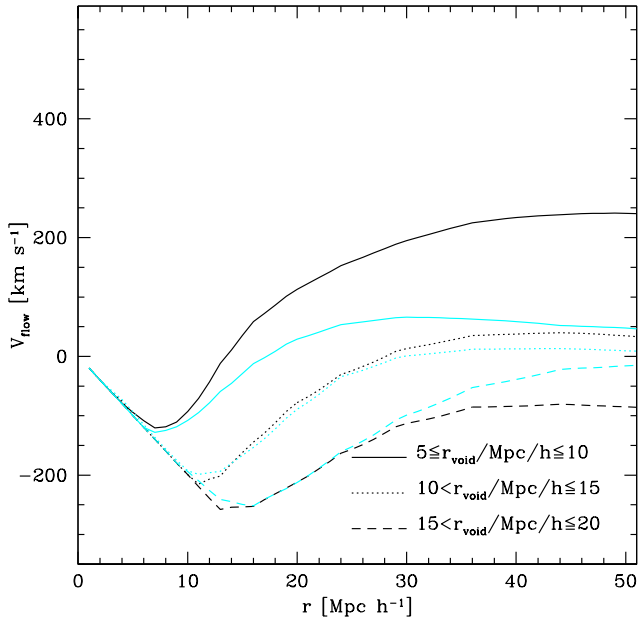}}
\caption{ Predicted velocity field amplitude as a function of  
distance to the void centre for the 2dFGRS (black lines) using the non-linear
velocity model.  Different line-types correspond to different void radii,
and the subsamples shown are the same as in figure \ref{fig:fn}. Predicted velocities for the mock 
catalogue are shown in gray lines for the same samples.
} 
\label{fig:fn2df}
\end{figure}

\begin{figure}
\epsfxsize=0.5\textwidth
\hspace*{-0.5cm} \centerline{\epsffile{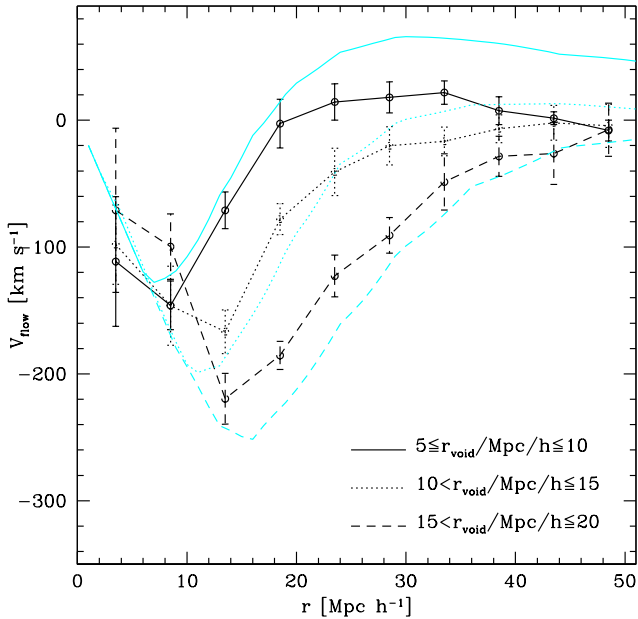}}
\caption{Outflow as a function of distance to the void 
centre derived from peculiar velocities in the mock catalogue (black lines). 
Different lines correspond to different ranges of void radius and are 
obtained from different subsamples 
(solid lines, $5\leq r_{void}/Mpc/h\leq 10$ and S121 , 
dotted lines, $10< r_{void}/Mpc/h\leq 15$ and S151, 
dashed lines, $15< r_{void}/Mpc/h\leq 20$ and S171). 
Gray lines show the velocity amplitude predicted by the outflow model using
the same samples and void radii as for the black lines} 
\label{fig:fn}
\end{figure}

In Figure \ref{fig:fn} we show the peculiar velocity outflow 
as a function of the distance to the void centre, derived from the mock 
peculiar velocities (black lines, the model results are shown in gray lines
for comparison). 
Different line types correspond 
to  different void radii and samples. 
As can be seen in the figure, 
the minimum velocity occurs at distances of about the void radius, so that
larger voids show a minimum 
velocity amplitude (maximum outflow) at greater distances 
from the void centre. The absolute velocity minimum is larger for greater
voids and the range of maximum outflow velocities goes from about $150 km/s$ 
for the smallest voids to more than $200 km/s$ for the largest voids.  
By comparing between measured and 
model outflow velocities (black and gray lines, respectively)
we can notice a qualitative agreement, specially for the larger voids.  Small
voids show some discrepancies at distances a few times the void radius;
this can be due to failure of the approximations assumed in the identification
of voids of small size, such as spherical symmetry.  

The maximum 
measured outflow velocities are successfully described by the model, as well as the distance 
to the void centre where the maximum outflow velocity occurs.  
Inspection of figures \ref{fig:fn2df} and \ref{fig:fn} shows that model
and measured velocities around
medium and large size voids (dotted lines and dashed lines, respectively) 
first reach a minimum value and then increase in value to 
eventually reach $V_{flow}=0$km/s. Notice that for the largest voids there are still
negative velocities at large distances. 
Small voids (solid lines) show positive velocities at distances of about two void radii,
and stay on positive values over a wide range of distances 
to finally return to $V_{flow}=0$ km/s. 
This behavior could be related with an infall of void shells
due to surrounding overdensities.
It is also related with the evolution of voids analised by Sheth and van de 
Weygaert (2004) in large scale numerical simulations, where 
they use two processes to describe the void evolution: void-in-void processes and void-in-cloud processes. 
The void-in-cloud refers to small voids embedded in dense regions, 
which show collapse instead of expansion.
Then, the infall of small void's shells can be related to the collapse in the void-in-cloud process.

\begin{figure}
\epsfxsize=0.5\textwidth
\hspace*{-0.5cm} \centerline{\epsffile{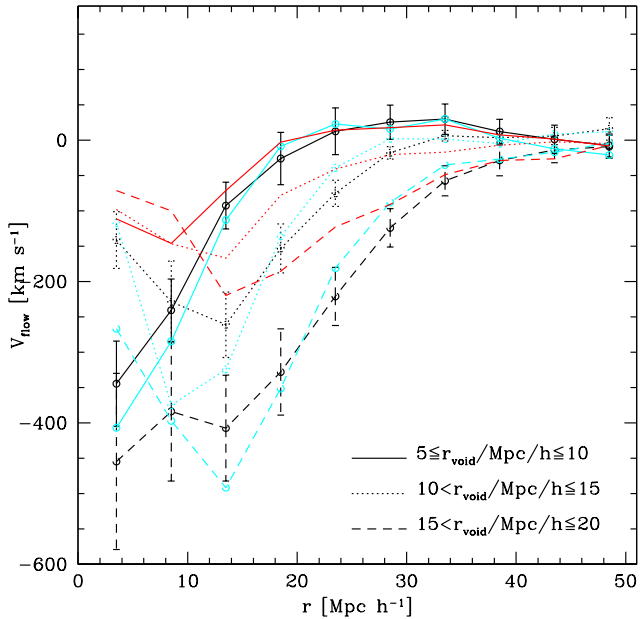}}
\caption{Velocity field amplitude as a function of the distance to the void centre
when including distance measurements errors. 
Dark-gray lines without error bars correspond to the measured outflow velocities
using peculiar velocity data with no errors; light-gray lines show the results
when including errors in the peculiar velocity; black lines show the results
when only using peculiar velocities (with errors) of 
galaxies within $45$ degrees of the line of sight from the
void centres.
Different line types correspond to 
different ranges of void radius and samples. 
} 
\label{fig:fner}
\end{figure}

In order to test the effect of distance measurement errors 
on the velocity field amplitude, we apply a random uncertainty proportional to 
$10\%$ of the redshift-space distance.
As expected, the main effect of distance uncertainties is to 
overestimate the peculiar velocity field amplitude specially in
the deepest samples.  The resulting velocity amplitudes are shown in light-gray lines in figure 
\ref{fig:fner} whereas in dark-gray lines we show 
the resulting velocity outflows when distance measurement errors are not included. 

As can be seen in figure \ref{fig:fner} there is a strong shift 
of outflow velocities to more negative values. For example for the greatest voids
(dashed lines), the maximum outflow velocity 
changes from 200 km/s to 400 km/s, a systematic increase of $100\%$ when including
distance measurement errors.    

The largest contribution from distance measurement errors to the systematic
offset in the outflow velocity comes from using radial peculiar velocities to infer
the outflow velocity when the angle subtended by the observer and the galaxy from
the void centre is close to $\pi/2$.  In this case, the component of the peculiar
velocity in the radial direction to the void centre is nearly null, and the deprojection
produces large spurious outflow velocities responsible for the systematic offset.  
Therefore, we repeat the analysis using only galaxies within $\pi/4$ of the line-of-sight 
(as seen from the void centres) and show the results in black lines in figure 
\ref{fig:fner}.  As can be seen the systematic offset is much less important in this case.

\subsubsection{Maximum outflow amplitude}
The results obtained in the previous subsections strongly suggest
that there is a relation between the the void size and
the distance where the maximum outflow occurs. The void radius seems to 
determine in a very simple way the distance of the maximum outflow velocity.
In order to characterise in a more quantitative way such a relation   
we study the contour lines of outflow velocity as a function of void
radius and distance to the void centre. 

\begin{figure}
\epsfxsize=0.5\textwidth
\hspace*{-0.5cm} \centerline{\epsffile{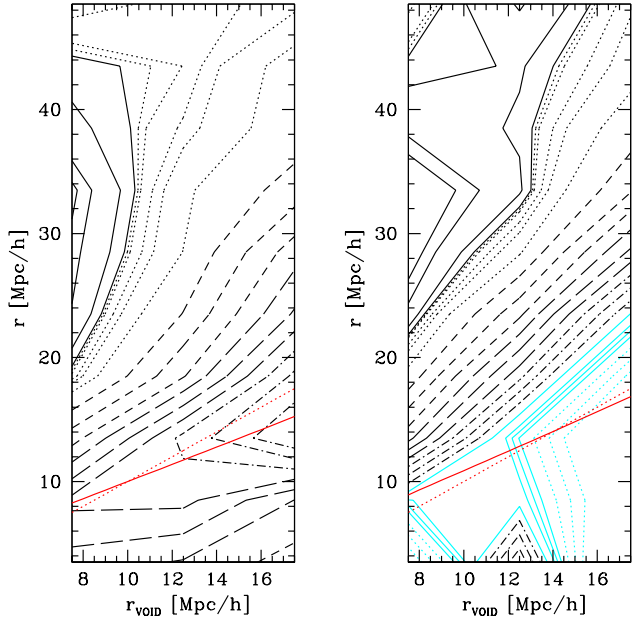}}
\caption{ Contour lines of outflow velocities as a function of void 
radius (x-axis) and distance to the void centre (y-axis) for the mock catalogue,
using peculiar velocities. 
The left panel shows the results from peculiar velocities with no errors;
the right panel shows the results when distance measurement errors are added.
Dot-dashed lines represent velocities lower than $-160$km/s ($-200$, $-180$, $-160$),
long-dashed lines represent velocities $-160<v/km/s<-100$ ($-140$, $-120$, $-100$),
short-dashed lines represent velocities $-100<v/km/s<-40$ (-$80$, $-60$, $-40$),
dotted lines represent velocities $-40<v/km/s<0$ ($-20$, $-10$, $-6$, $-2$, $-1$),
 and 
solid lines represent positive velocities ($0$, $5$, $15$, $20$).
} 
\label{fig:flow2d}
\end{figure}

The contour lines of outflow velocity 
directly measured from the mock peculiar velocities are plotted in the left panel of Figure 
\ref{fig:flow2d}, where the x-axis represents the void radius and the y-axis
the distance to the void centre.  The solid lines in figure \ref{fig:flow2d},
which represent positive velocities (i.e. galaxy infall to the void centre), 
are only found for small voids ($r_{void}<11 Mpc/h$) at relatively large 
distances compared to the void size. It is easily seen how large outflow velocities
($v_{out}>100 km/s$) are found around the largest voids
(with radii $r_{void}> 14 Mpc/h$) at distances in the range $ 10 <r Mpc/h< 20$. 
Note that this range is approximately centred at the average void radius.
The right panel shows the results when adding distance measurement errors
to the mock peculiar velocities.  As can be seen, there is also a clear
relation between void radius and the distance corresponding to the maximum
outflow velocity; however, as a consequence of large velocity errors,
the resulting outflow velocities are much higher in this case.

In order to characterise the relation between the void radius an the distance at 
which the minimum velocity occurs, we find that a linear relation 
fits the results:
\begin{equation} 
r(v_{min})=a_l r_{void}+b_l,
\label{ec:lin}
\end{equation} 
where $a_l=0.7$ and $b_l=2$. This expression is represented by a gray dotted line in 
both panels of figure \ref{fig:flow2d}. Although this is the simplest way to describe 
the relation $r(v_{min})$-$r_{void}$ we also fit an exponential relation:
\begin{equation} 
r(v_{min})=exp(r_{void}/a)^{b}
\label{ec:exp}
\end{equation} 
which is shown in a gray solid line in figure \ref{fig:flow2d}, 
with best fitting parameters $a=0.62$ and $b=0.30$ (no errors) and
$a=0.55$ and $b=0.33$ for the results when including distance measurement errors.
This exponential expression can also fit the distance where 
the velocities reach $0km/s$; in this case, the best fitting
parameters are $a=1$ and $b=0.5$ (when not considering errors). 
The parameter values obtained from peculiar velocities with no errors are similar to those 
obtained by fitting the same expression to semianalytic 
galaxies (Padilla et al, 2005).  As can be seen in figure \ref{fig:flow2d}, 
the linear and exponential functions are very similar, 
and both can be used to give an approximate relation between radius and 
maximum outflow amplitude distance.

\begin{figure}
\epsfxsize=0.5\textwidth
\hspace*{-0.5cm} \centerline{\epsffile{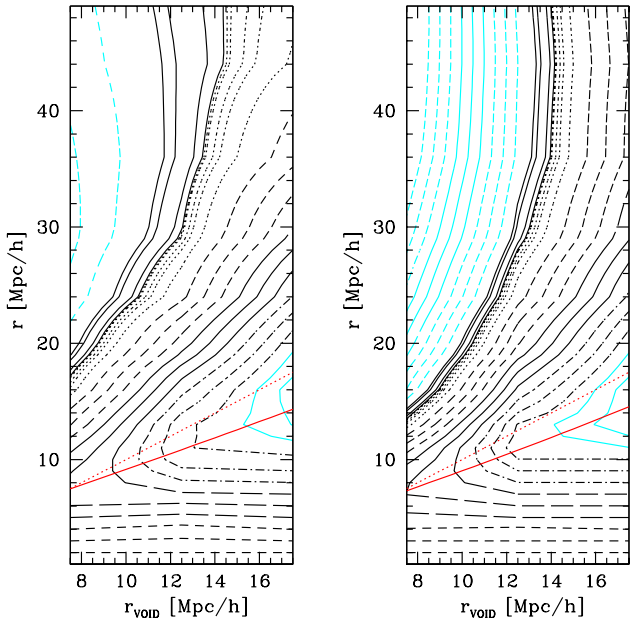}}
\caption{ Contour lines of peculiar velocity outflows derived from the theoretical 
model,
as a function of void radius (x-axis) and distance to the void centre (y-axis), 
for the mock and 2dFGRS catalogues (left and right panels respectively).
Gray solid lines represent 
velocities greater than $-340$ km/s ($-340$, $-320$, $-300$, $-280$),
gray dotted lines $-280<v/km/s<-200$ ($-260$, $-240$, $-210$),
black dot-dashed lines $-220<v/km/s<-140$ ($-200$, $-180$, $-160$),
black long-dashed lines $-160<v/km/s<-100$ ($-140$, $-120$, $-100$),
black short-dashed lines $-100<v/km/s<-40$ ($-80$, $-60$, $-40$),
black dotted lines $-40<v/km/s<0$ ($-20$, $-10$, $-6$, $-2$, $-1$),
black solid lines represents positive velocities larger than 
$10$ km/s ($0$, $5$, $15$, $20$), 
cyan short-dashed lines $20<v/km/s<100$ ($40$, $60$, $80$),
cyan long-dashed lines $160<v/km/s<100$ ($100$, $120$, $140$), and
cyan dot-dashed lines $ v>160 km/s$ ($160$, $180$, $200$).
} 
\label{fig:flow2dteo}
\end{figure}

We analise the dependence of the maximum outflow distance on 
void radius predicted by the non-linear outflow model, in this
case for both mock and observational catalogues.
The aim  in doing this is to compare the results from both catalogues, and 
to test the quality and efficiency of the model to describe the velocity field.  
We produce the velocity contour lines predicted by the model 
and show them in figure \ref{fig:flow2dteo}, where the left panel corresponds to 
the mock catalogue and the right panel corresponds to the 2dFGRS.

If we compare both panels of figure \ref{fig:flow2dteo} we note
that from a qualitative point the results from mock and theory are quite similar.
In a more quantitative analysis,
we find that the maximum velocities around large voids ($r_{void}>
14$h$^{-1}$Mpc) occur at scales in the range 
$ 10 <r$h$^{-1}$Mpc$< 20$ in both cases. The results shown in this figure are
also in agreement 
with those found from the direct measurement of the peculiar velocity field shown 
in figure \ref{fig:flow2d}.

We fit the maximum model outflow velocities using equations \ref{ec:lin} and \ref{ec:exp}.
The results from the mock catalogue are best fitted using the exponential expression
with $a=0.90$ and $b=0.33$; the results from the 2dFGRS theory predictions are fitted
by $a=1.05$ and $b=0.35$, also for the exponential fit.  Note the similarity
between the results from mock and real data.

\subsubsection{Velocity dispersion in void shells}
In order to fully characterise the dynamics of galaxies in void shells,   
we also compute the peculiar velocity dispersion 
in the directions parallel and perpendicular to the void shell.
Again, due to the lack of peculiar velocity information in the 2dFGRS survey,
all the analyses in this subsection are restricted to
the mock catalogues.

As we only have the line-of-sight projection
of the peculiar velocity, 
the best signal for velocities perpendicular to void shell (in the
void radial direction) comes 
from galaxies close in projection to the void centre, whereas  
for the velocity parallel to the void shell, this comes from galaxies
away from the void centre, in projection. 
Taking this into account we use galaxies within $45$ degrees of
the line of sight from the void centre when calculating the dispersion in 
the direction perpendicular to the void shell,
and only galaxies at more than 45 degrees with respect to 
the line of sight from the void centre when calculating the dispersion in 
the direction parallel to the void shell.

\begin{figure}
\epsfxsize=0.5\textwidth
\hspace*{-0.5cm} \centerline{\epsffile{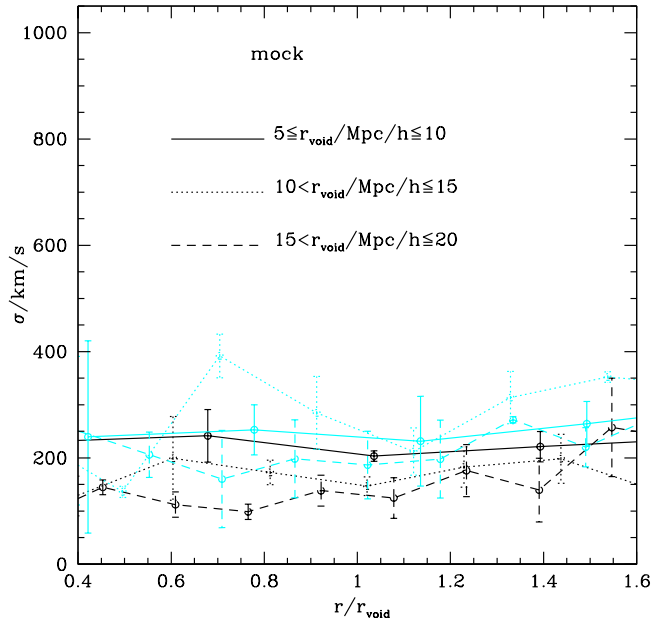}}
\caption{ 
Galaxy peculiar velocity dispersion relative to the void shell as a function
of normalised distance to the void centre.
Black lines show the dispersion in the direction perpendicular to the 
void shells and gray (cyan) lines indicate dispersion in the direction parallel
to the void shells. Different line-types correspond to different void radii 
and galaxy samples:
solid lines are for $5\leq r_{void}/Mpc/h\leq 10$ and S121 , 
dotted lines are for $10< r_{void}/Mpc/h\leq 15$ and S151, and 
dashed lines are for  $15< r_{void}/Mpc/h\leq 20$ and S171. 
} 
\label{fig:sig}
\end{figure}

Figure  \ref{fig:sig} shows the relative velocity dispersion as a function of distance
to the void centre in the mock catalogue. Black lines show the dispersion about 
the mean motion in the radial direction, and gray lines show the dispersion in the  
direction parallel to the void walls. Solid lines correspond to small 
voids from sample S101, dotted lines correspond to intermediate size voids from sample 
S151 and dashed lines correspond to large voids from sample S171.
Errors are calculated considering both Poisson and jackknife errors, added in quadrature.  

It can be seen in figure \ref{fig:sig}
that the dispersion is not very dependent on 
distance to the void centre, specially for the smallest voids.
The plot also shows that parallel velocities are always 
higher than radial velocities indicating that
galaxies tend to move faster in the direction of void shell.
It can also be seen that the smallest voids 
show larger values of velocity dispersion than the largest voids.
All these results are consistent with those obtained using semianalytic galaxies 
in a full simulation box
(Padilla, Ceccarelli \& Lambas, 2005).

\begin{figure}
\epsfxsize=0.5\textwidth
\hspace*{-0.5cm} \centerline{\epsffile{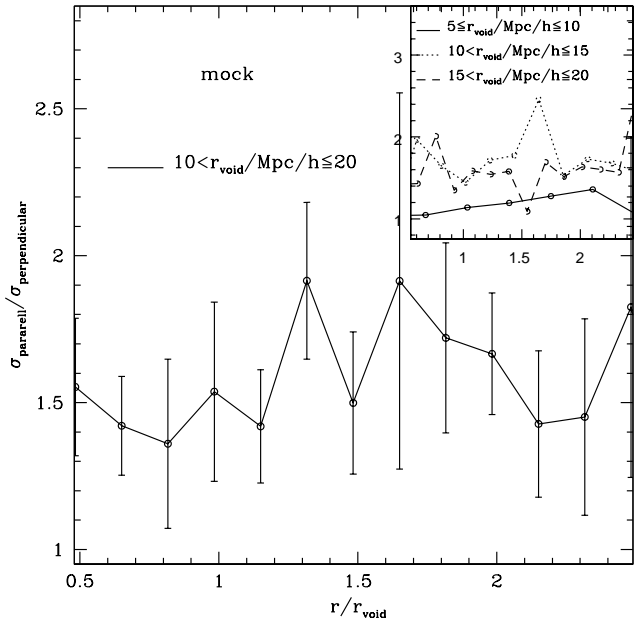}}
\caption{$\sigma_{parallel}/\sigma_{perpendicular}$
as a function
of normalised distance to the void centre for $10< r_{void}/Mpc/h\leq 20$;
inner panel:
$\sigma_{parallel}/\sigma_{perpendicular}$
as a function
of normalised distance to the void centre.
Different line-types correspond to different void radii 
and galaxy sample:
solid lines correspond to $5\leq r_{void}/Mpc/h\leq 10$ and S121 , 
dotted lines to $10< r_{void}/Mpc/h\leq 15$ and S151 and 
dashed lines to $15< r_{void}/Mpc/h\leq 20$ and S171. 
} 
\label{fig:sigrel}
\end{figure}

We also show in figure \ref{fig:sigrel} the ratio between the
velocity dispersion in the directions parallel and perpendicular to the void shell.
In order to improve the statistics we have used a 
combined sample of intermediate and larger voids $10 < r_{void} \leq 20$.
For this sample the ratio between parallel and perpendicular 
dispersion is shown in figure \ref{fig:sigrel}. 
In the inner panel are shown the ratios  
$\sigma_{parallel}/\sigma_{perpendicular}$
obtained from the same samples analysed in figure \ref{fig:sig} 

As it can be seen in the large panel of figure \ref{fig:sigrel}, 
the ratio of parallel to perpendicular dispersion is nearly constant at 
$\sigma_{parallel}/\sigma_{perpendicular} \sim 1.5$ for  distances 
to the void centre in the range $0.5<r/r_{void}<2.5$.  Note that these values
are comparable to those found in numerical simulations 
by Padilla, Ceccarelli \& Lambas (2005).

\begin{figure}
\epsfxsize=0.5\textwidth
\hspace*{-0.5cm} \centerline{\epsffile{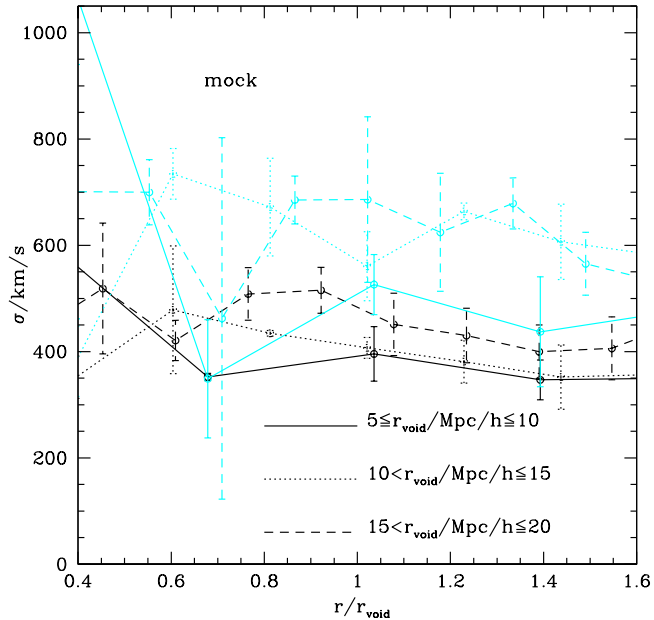}}
\caption{
Galaxy peculiar velocity dispersion relative to the void shell as a function
of normalised distance to the void centre affected by distance measurement errors.
Black lines indicate the dispersion in the direction perpendicular to the 
void shell, and gray (cyan) lines indicate the dispersion in the direction parallel
to the void shell. Different line-types correspond to different void radii 
and galaxy samples as in figure \ref{fig:sig}.
} 
\label{fig:siger}
\end{figure}

We add distance measurement errors to the galaxy peculiar velocities
proportional to the distance to the observer (see for instance Padilla \& Lambas, 1999).
The velocity dispersions obtained using peculiar velocities affected by distance
measurement errors are shown in figure 
\ref{fig:siger}, where black lines show the velocity dispersion  
along the direction perpendicular to the void shell and gray
lines show the dispersion parallel to the void shell. 
Solid lines correspond to small voids from sample S101, 
dotted lines correspond to intermediate voids from sample S151 and 
dashed lines correspond to large voids from sample S171 as in figure
\ref{fig:sig}.

Inspection of figure \ref{fig:siger} shows that velocity dispersions are higher when 
peculiar velocity errors are included.
We can also see that this effect is more important for the largest voids. 
The reason behind this systematic offset 
is that  we use samples that reach higher redshifts for larger voids. 
In spite of this,  a qualitative comparison of figures 
\ref{fig:sig} and \ref{fig:siger} 
shows that the behavior of the velocity dispersion parallel and perpendicular 
to the void shells are similar in both plots and that it can still be detected 
that galaxies move faster along the void walls, or alternatively, filaments.

\section{CONCLUSIONS}

In this work we have studied statistical properties of 
the dynamics of voids and galaxies populating void shells.
We were specially interested in detecting and characterising 
the expansion of voids, and in order to do so, we studied the redshift space 
distortions detected in the galaxy-void
cross correlation function measured in the 
directions parallel and perpendicular to the 
line-of-sight. We also analysed the amplitude of the peculiar 
velocity outflows obtained from the peculiar velocities in mock catalogues
and from non-linear theory approximation (Croft, Dalton \& Efstathiou, 1999).
In addition, we analysed the velocity dispersion in the  directions parallel and 
perpendicular to the void shell.   
In all our statistical analyses we test the effect of peculiar velocity errors 
in our results.  We summarise our results as follows:

             \begin{itemize}
             \item {
We applied a void finding algorithm, 
which is capable of accurately identifying voids in galaxy catalogues 
affected by angular and radial selection functions, 
to volume limited samples of mock and 
observational 2dFGRS catalogues.}
             \item {The redshift space distortion of galaxies around voids shows 
elongations along the line of sight. This can be interpreted as evidence of an outflow
motion in both mock and observational data.}
            \item {Outflow velocities from the mock catalogue (without errors) are
consistent with the non-linear theory predictions based on measurements of the 
correlation functions in the mock catalogue. }
	    \item {By applying the non-linear outflow model to 2dFGRS data, we find that
larger voids would be characterised by larger maximum outflow
velocities: $ 110$km/s,
$210$km/s and $270$km/s for $<r_{void}>=7.5$, $12.5$ and $17.5$h$^{-1}$Mpc
respectively; assuming a galaxy bias $b=1$.  We find that 
out to a distance of the order of the void radius.
The results from the application of the non-linear outflow model
to observational data are consistent with results
from the mock catalogues  
.}
	    \end{itemize}

Our analysis of mock catalogues indicates that several additional results on the dynamics
around voids could be derived from the 2dFGRS survey if peculiar velocity data was available
as well, or at least for a few tens of thousands of galaxies sampling a large enough
volume so that a reasonable number of voids could be found,
	    \begin{itemize}
            \item {By studying the mock catalogues, we are able to perform a 
direct measurement of galaxy outflows using peculiar velocities.  Outflows of 
about 200 km/s are measured, in accordance
with the results from numerical simulations from Padilla et al.(2005).  
}
            \item {
When distance measurements errors are included in the mock galaxies,
the estimated outflow velocity is biased to higher values, which could be up to 400 km/s for the larger voids.
}
            \item {Measurements of the velocity dispersion in shells around voids
in the mock catalogues
show that these are larger in the direction parallel to the void walls, 
in agreement with results from the full numerical simulation (Padilla, Ceccarelli
\& Lambas, 2005). This result indicates
that galaxies tend to move preferentially in the direction parallel to
void shells, that is, along large-scale structure filaments and walls, and that
a peculiar velocity catalogue with sufficient data could allow this measurement
to be done, constituting a new test to the hierarchical clustering paradigm.
}
             \end{itemize}

\section*{Acknowledgments}
This work has been partially supported by Consejo de Investigaciones 
Cient\'{\i}ficas y T\'ecnicas de la Rep\'ublica Argentina (CONICET), the
Secretar\'{\i}a de Ciencia y T\'ecnica de la Universidad Nacional de C\'ordoba
(SeCyT), Fundaci\'on Antorchas, Argentina and Agencia C\'ordoba Ciencia. 
This work was supported in part by the ESO-Chile Joint Committee,
NP was supported by a Proyecto Fondecyt Postdoctoral no. 3040038.
LC and CV were partially supported by the Latin-american and European Network 
for Astrophysics and Cosmology (LENAC) Project.

\end{document}